# Introducing Shape Prior Module in Diffusion Model for Medical Image Segmentation

1st Zhiqing Zhang
*Shenzhen Institute of Advanced Technology*
*University of Chinese Academy of Sciences*
Shenzhen, China
zq.zhang3@siat.ac.cn

2nd Guojia Fan*
*College Of Information Science and Engineering*
*Northeastern University (of Affiliation)*
Shenzhen, China
gj.fan@siat.ac.cn

3rd Tianyong Liu
College Of Geosciences
*Northeast Petroleum University*
Shenzhen, China
ty.liu1@siat.ac.cn

4th Nan Li
*Department of Computer Science*
City University of Hong Kong
Shenzhen, China
nanli69-c@my.cityu.edu.hk

5th Yuyang Liu
*Shenyang Institute of Automation*
*University of Chinese Academy of Sciences*
Shenzhen, China
sunshineliuyuyang@gmail.com

6th Ziyu Liu
Faculty of Science and Engineering *Chinese Academy of Sciences*
*University of Nottingham Ningbo China*
Ningbo China
zy.liu4@siat.ac.cn

7th Canwei Dong
College Of Information Science And Engineering
*Northeastern University(of Affiliation)*
Shenzhen,China
cw.dong@siat.ac.cn

8th Shoujun Zhou
*Shenzhen Institute of Advanced Technology*
*University of Chinese Academy of Sciences*
Shenzhen, China
sj.zhou@siat.ac.cn

*Abstract*—Medical image segmentation is critical for diagnosing and treating spinal disorders. However, the presence of high noise, ambiguity, and uncertainty makes this task highly challenging. Factors such as unclear anatomical boundaries, inter-class similarities, and irrational annotations contribute to this challenge. Achieving both accurate and diverse segmentation templates is essential to support radiologists in clinical practice. In recent years, denoising diffusion probabilistic modeling (DDPM) has emerged as a prominent research topic in computer vision. It has demonstrated effectiveness in various vision tasks, including image deblurring, super-resolution, anomaly detection, and even semantic representation generation at the pixel level. Despite the robustness of existing diffusion models in visual generation tasks, they still struggle with discrete masks and their various effects. To address the need for accurate and diverse spine medical image segmentation templates, we propose an end-to-end framework called VerseDiff-UNet, which leverages the denoising diffusion probabilistic model (DDPM). Our approach integrates the diffusion model into a standard U-shaped architecture. At each step, we combine the noise-added image with the labeled mask to guide the diffusion direction accurately towards the target region. Furthermore, to capture specific anatomical a priori information in medical images, we incorporate a shape a priori module. This module efficiently extracts structural semantic information from the input spine images. We evaluate our method on a single dataset of spine images acquired through X-ray imaging. Our results demonstrate that VerseDiff-UNet significantly outperforms other state-of-the-art methods in terms of accuracy while preserving the natural features and variations of anatomy. The generality and validity of our proposed model are confirmed through extensive experiments. By enabling more precise segmentation of anatomical structures, our framework has the potential to facilitate accurate diagnosis and treatment of medical conditions by enabling more precise segmentation of anatomical structures.

*Keywords—Diffusion probabilistic modelling (DDPM), Vertebral segmentation*

## I. INTRODUCTION

Spinal surgical diseases are prevalent and have a significant impact on patients' quality of life [1,2]. Accurate segmentation of spine images, including the vertebrae and intervertebral discs (IVDs), is crucial for various orthopaedic applications such as evaluation, diagnosis, surgical planning, and image-guided interventions. For example, internal fixation with spinal pedicle screws has been widely used over the years in surgery for spinal degeneration, trauma, deformity, inflammation, tuberculosis, and tumours due to its ability to achieve three-column fixation, provide immediate and firm fixation of the spine, strong orthopaedic power, and the ability to substantially increase fusion rates [3]. Poor and incorrect nail placement can lead to fixation failure and potentially serious neurovascular injury, so improving the accuracy of pedicle screw placement has become the key to the success of pedicle screw internal fixation [4]. However, due to the limited field of view of the surgeon and the difficulty in identifying anatomical landmarks, it is difficult to achieve good nail placement in most cases. X-ray-assisted nail placement is a technique that assists the surgeon in determining the nail placement point by displaying the position of the screws in the two-dimensional image through intraoperative X-ray fluoroscopy, which is widely used in the clinic. However, this technique has limitations due to the lack of three-dimensional images and the presence of ghosting and noise in the X-ray images, making manual determination of screw position challenging [5]. Another minimally invasive interventional therapy for the treatment of lumbar pain caused by lumbar disc herniation is lumbar radiofrequency ablation (RFA) [6]. But there are also some inherent defects of this treatment method, such as better contrast of the vertebral body in CT images, but MR images in turn have high contrast for intervertebral discs. Therefore, fusing the multimodal information of CT images and magnetic resonance (MR) images and accurately segmenting specific spinal structural regions will effectively promote the development of this medical technology.

However, due to the fact that the current spine medical images not only have the characteristics of fuzzy images, uneven grey scale distribution, multiple noises, and low image contrast, but also because the spine structure is composed of a series of vertebrae with similar structural shapes and different species, the current parsing based on these images is still faced with problems such as interclass similarity, intra-group variation [7], spatial bias [8], and even high computational and memory costs of high-dimensional spine medical images, all of which lead to problems such as inter-class similarity, intra-group variations [7], spatial offset [8], and even high-dimensional medical images of the spine. cost, all of which make medical images of the spine more difficult to process compared to other types of medical images. In addition, when common algorithms for general image processing are applied to spine medical images, the results are not satisfactory, and it is difficult to form a unified and effective processing algorithm. When designing medical image processing methods for the spine, it is often necessary to have a strong focus on different types of images or different types of parts that have their own specific processing methods to improve and optimize the traditional image processing methods, or propose some novel and unique processing methods. At present, the common research on spine images is divided into image pre-processing, intervertebral disc localization and segmentation, and vertebral block localization and segmentation. The localization and detection of the spine in CT, MR, and even X-ray images are of great significance for the diagnosis and treatment of various spinal diseases.

*A. Related Works*

Early work on vertebral localisation and detection relied on mathematical models and traditional manual features such as deformable models [9,10], custom filters[11], atlas-based [12,13] methods and machine learning [3].

For example, in 2012 Prabhu [9] et al. proposed a method to list all vertebral horizontal inclinations using an active contour model and morphological operators. In 2017 Ibragimov [10] et al. combined landmark detection and deformation modelling into a supervised multi-energy segmentation framework for segmenting fractured lumbar vertebrae from CT images. The framework employs the Laplace shape editing theory introduced in the field of computer graphics, thus avoiding the limitations of statistical shape modelling.

An automated system for extracting desired anatomical features using custom filters was proposed by Anitha et al [11] in 2014. The combination of these filters automatically extracts anatomical features based on the desired vertebral endplates and eliminates the human intervention involved in Cobb angle quantification.

In 2015 D. Forsberg [12] used label fusion to combine the labels of deformation atlases in order to obtain the final segmentation of the target dataset. Ultimately a DICE metric of 0.94 on average was achieved on the target dataset. In addition, Fallah et al.'s [13] method combines an atlas-based approach with machine learning to achieve DSC with 92.5% VB segmentation of vertebrae on fat-water MR images from a subset of a large cohort dataset, using a hierarchical random forest classifier (HRF) and a hierarchical conditional random field (HCRF) based on a set of local and contextual features on a multiscale scale.

In terms of machine learning, in 2012 Glocker et al [3,14] used regression forests and probabilistic graphical models to detect vertebrae in CT scans of arbitrary fields of view and subsequently proposed a supervised classification forest-based method for vertebrae localisation and identification. Subsequently, Chen et al [15] proposed a machine-learning-based approach for localisation and segmentation of intervertebral discs on volumetric MR images. Segmentation of intervertebral discs was achieved by locating the centre of each disc in a data-driven manner to classify the image voxels around the disc centre and ultimately obtained disc segmentation results with an average dice coefficient of 85%-88%.

However, due to the limited modelling ability of hand-constructed features or the low robustness of machine learning-based segmentation methods, features need to be extracted manually thus missing task specificity and limiting the performance of segmentation. With the advancement of deep learning techniques, several studies have successfully applied neural network-based models, including the classical convolutional neural networks (CNNs) [16,17] and the recently popular visual transformers (ViTs) [18-20] for medical image segmentation tasks.

Traditional medical segmentation algorithms usually use an encoder-decoder structure [21] and incorporate jump connections to enable the decoder to reuse the features extracted by the encoder. However, since its structure is based on convolutional neural networks, it cannot extract global features efficiently. On the contrary, the structure of the Transformer has significantly improved the performance of global feature modelling, but the corresponding dynamic extraction of global features not only leads to an increase in the amount of computation but also results in a decline in compatibility and difficulty in convergence of iterations [22-25].

Recently, the Diffusion Probabilistic Model (DPM) [26] has gained popularity as a powerful generative model gained popularity as a powerful generative model capable of producing high-quality and diverse images [27,28]. Inspired by its success, several researchers have applied DPM initially to the field of medical image segmentation [29-33]. Most of these methods are based on the classical decoding and coding structure of UNet, on which some researchers introduced the transformer architecture in order to extract deeper semantic segmentation features from the original image, but often ignored the difficulty and geometrically increasing computational effort to design the transformer segmentation architecture to satisfy the compatibility.

In order to be able to improve the ability of the diffusion-based segmentation model to capture prior knowledge of spine morphology, we introduced a morphology prior module to significantly improve the segmentation performance of the DPM-based model [26]. We validate the effectiveness and generalisation of the model on three different modalities of medical images, namely, spine image in a single MR, spine image in a single X-ray and spine image in a mixture of MR and CT.

In this paper, we propose a novel approach for general medical image segmentation using the Deep Probabilistic Model (DPM) framework..

## II. METHOD

In this section, we briefly introduce our proposed diffusion modelling framework. Figure 1 shows the training phase, sampling phase of our proposed VerseDiff-UNet. Unlike traditional medical image segmentation methods that directly input the original image data to predict the corresponding segmentation label maps, the diffusion model learns the denoising process using the original image and the segmentation label maps with noise as inputs.

### A. Label and time Embedding

we add successive $t$ step noise $\varepsilon$ for the converted multi-channel labels, called the diffusion forward process.

$$x_t = \sqrt{\overline{\alpha_t}}x_0 + \sqrt{1-\overline{\alpha_t}}\epsilon$$

After getting the labelmap $x_t$ with t step noise, our target is to predict the clear label map $x_0$ based on $x_t$ and the raw image data by the Denoising Module.

### B. Denoising Module

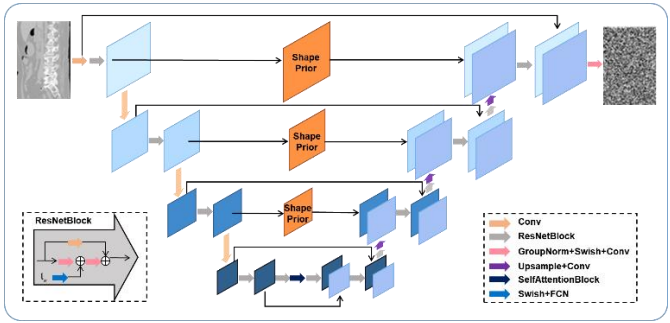

Fig. 1. Diagram of the proposed overall network architecture

As shown in Fig. 1, The Unet in our proposed *VerseDiff-UNet* also contains two parts, an encoder and a decoder. First, given the original image data I $\in R^{N\times D\times W\times H}$, where N is the number of modal images, I and the noisy one-hot label $x_t$ channel are cascaded into the encoder of *VerseDiff-UNet* to obtain the multiscale feature $I_E$. The information input to the network is not only extracted through the traditional downsampling and upsampling structure, but also through the per-layer shape prior science The information is extracted from the $S_f$ module and summed with the corresponding decoder features $I_D$ at the same level to obtain the multiscale fusion features $I_f$. Since the same level of encoder features $I_E$ and decoder features $I_D$ contain the same number of features with the same size, the skipped original features can be further enhanced by interacting with the global shape prior, thus facilitating the generation of features with discriminative shape representations and global context, which thus helps the model to better extract the vertebral anatomical prior features of the original medical image ultimately leading to the prediction result $\bar{x}_0$. For the task of medical image segmentation, our network directly generates the predicted mask as the output.

Among them, The step-index t is integrated with the added embedding and decoder features. In each of these, it is embedded using a shared learned look-up table.

$$\bar{x}_0 = Verdiffu(cat(I, x_t), t, I_f)$$

### C. Training and Sampling Architecture

As shown in Fig. 2, the main architecture of *VerseDiff-UNet* is a Unet-based coder-decoder. The coder-decoder consists of three corresponding downsampling and upsampling modules each.

Diffusion models perform variational inference on a Markovian process using time steps to learn the training data distribution p($x_0$). In the training phase, for every step, the anatomical structure is induced by adding the input image to the noisy segmentation mask.

Specifically, the training phase inputs model pairs of original spine images with manually labelled masks, and the masks are progressively noise-added. The framework consists of a forward and a reverse process, which is also called training and sampling process. During the forward process for each timestep in T, Gaussian noise is added to the image ~ p() until the image becomes an isotropic Gaussian. This forward noising process is denoted as

$$q(x_t|x_{t-1}) = \mathcal{N}(x_t; \sqrt{\alpha_t}x_{t-1}, (1-\alpha_t)I)$$

where $(x_0, x_1, x_2 \cdots x_T)$ denotes the $T$-step in the Markov chain, $\alpha$ is the noise scheduler that controls the variance of the noise. In the inverse process, an incremental denoising sequence is built using Markov chain with Unet network structure to obtain clean images. Sampling n times with different Gaussian noise, n different plausible masks are generated. The parameters:

$$p(x_{t-1}|x_t) = \mathcal{N}(x_{t-1}:\mu_\theta(x_t,t), \sum(x_t,t))$$

Which are obtained by minimizing the KL-divergence between the forward and the reverse distribution for all timesteps.

Identify applicable funding agency here. If none, delete this text box.

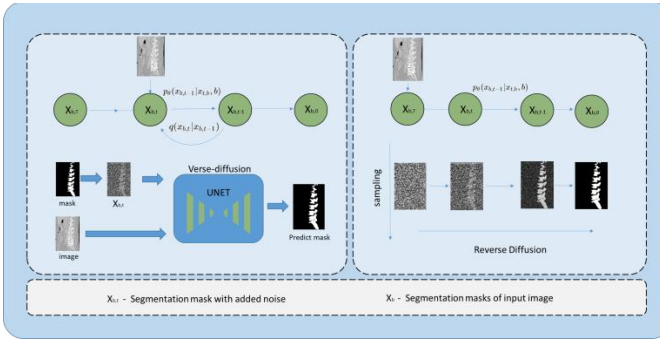

Fig. 2. The graphical model of a) sampling and b) training procedure

The loss can be represented as:

$$\mathcal{L} = E_{x_0,\epsilon,t}[||\epsilon - \epsilon_\theta(\sqrt{\hat{a}_t}x_0 + \sqrt{1-\hat{a}_t}\epsilon, I_i, t)||^2]$$

The iteration number is sampled from a uniform distribution and $\varepsilon$ from a Gaussian distribution.

## III. EXPERIMENTS AND RESULTS

### A. Dataset

The dataset consists of 609 spinal anterior-posterior x-ray images. The landmarks were provided by two professional doctors in London Health Sciences Center. Each vertebra was located by four landmarks with respect to four corners. All the image sizes are 250x750.In addition, since the noise artefacts in this dataset are extremely serious, and some of the data are even mixed with medical clinical instruments on the spine, in order to further test the segmentation ability of the proposed model under this special condition, we only target the spine region, and perform the 0/1 binary transformation of the region for the original labelling. Please note that although this operation method may cause the loss of the labelling region for some specific parts of the spine, the final experimental results effectively test the model's ability to segment the spine image under occlusion in this medical imaging condition..

### B. Data augmentation

We use dynamic data augmentation during training to avoid network overfitting and increase the robustness of our model.

Three types of data augmentation are included:

- Randomly rotating the image from -30° to 30° to simulate the rotation variance;
- Randomly shifting the image by 1-5% to simulate the shift variance;
- Random elastic deformation, and random contrast adjustment to improve the generalization of the model. To ensure all the methods used the same augmented training data, we used the same random state (seed = 35) for all methods when conducting data augmentation.

### C. Training and sampling

All the networks were implemented by Pytorch and codes ran in a server with an RTX 3090 GPU unless otherwise specified. Dice similarity coefficient (DSC), and IOU were used as the quantitative metrics to evaluate the segmentation performance. Both metrics were calculated for an individual object in the original image space, and then averaged across all subjects. The training and sampling/test sets of each kind of dataset were divided into 7:3 and comparative segmentation experiments were performed.

## IV. RESULTS AND ANALYSIS

To evaluate the segmentation performance of our proposed method, we perform comparative experiments with other state-of-the-art methods on publicly available CT datasets to demonstrate the effectiveness of different loss and estimation objectives. As shown in the Table 1 , we use Dice followed by IOU as the evaluation metric. Best results are denoted as bold. Dice similarity coefficient (DSC), IOU were used as the quantitative metrics to evaluate the segmentation performance.Tabl

As this dataset presents high noise, high artefacts and other features, while the target spine region presents a very high similarity with the background pixels in some images. For example, the swin-transfomer, which is sensitive to relative position information, cannot extract effective feature information of the spine region in the actual training process, the convergence iteration speed is too slow and ultimately cannot learn the semantic information of the target region. As for the current popular SAM model, as shown by the red circle in Fig. 3, SAM directly treats them as one in this case and loses the ability of semantic segmentation in this case, which in turn loses the stability of segmentation and reduces the segmentation ability in the overall dataset. In contrast, our proposed model is able to extract effective spine features in this case compared to the original mask, although it also loses some of the detailed features.

In addition, dividing the comparison shown in the red circle, FCN, Deeplabv3 loses the ability to capture the features of the spine region under the coverage of the external instrumentation, and the only part of the spine inferred is the edge of the region uncovered by the trailing end of the external instrumentation when compared with the original image. In contrast, our proposed model is still able to specifically and directly infer a given vertebra. According to our analysis, when the model learns the spine features during the noise addition process, it only learns the specific features in the spine region and treats the other parts of the region as noise, so in the reverse denoising stage, all the pixels above the spine layer are removed as noise and thus the target feature region containing only the spine is segmented smoothly under this counter-intuitive condition with masking.

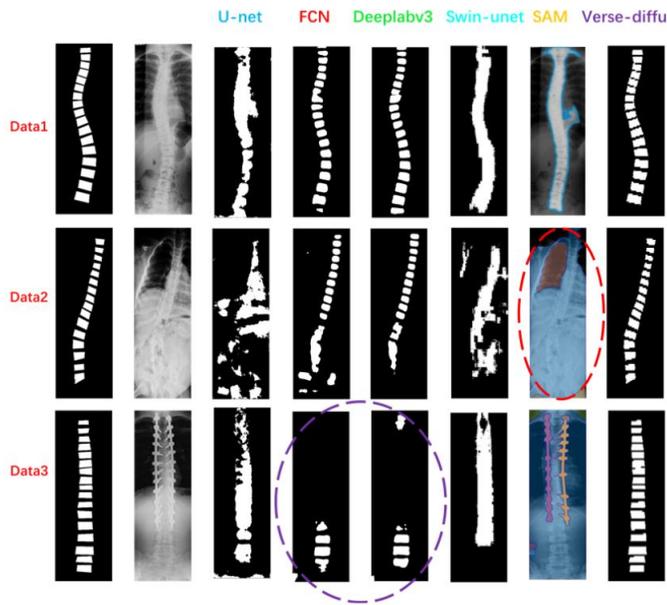

Fig. 3. Visual comparison of several generic medical image segmentation methods on the dataset. We randomly selected three photos as to show the results of different segmentation methods.

TABLE I. THE COMPARISON OF VERSE-DIFFUNET WITH SOTA SEGMENTATION METHODS.

| X ray-dataset | Evaluation metrics | |
|---|---|---|
| | Dice | IOU |
| Unet | 0.6078 | 0.4375 |
| Fcn | 0.6533 | 0.5266 |
| Deeplabv3-resnet50 | 0.7244 | 0.5791 |
| Swin-transfomer | 0.6893 | 0.5313 |
| SAM | 0.6789 | 0.5256 |
| Verse-diffu | **0.7865** | **0.6765** |

ACKNOWLEDGMENT

This work was supported in part by the National Key R&D Project of China (2018YFA0704102 and 2018YFA0704104), in part by the National Natural Science Foundation of China (No. 81827805), in part by Natural Science Foundation of Guangdong Province (No. 2023A1515010673), and in part by Shenzhen Technology Innovation Commission (No. JCYJ20200109114610201, and JSGG20220831110400001), in part by the Shenzhen Engineering Laboratory for Diagnosis & Treatment Key Technologies of Interventional Surgical Robots (XMHT20220104009).

## Authors' background

| Name | Prefix | Research Field | Email | Personal website |
|---|---|---|---|---|
| Zhiqing Zhang | PhD Candidate | Medical image processing, Diffusion model | zq.zhang3@siat.ac.cn | |
| Guojia Fan | Master Student | Medical image processing | gj.fan@siat.ac.cn | |
| Tianyong Liu | Master Student | Medical image processing | ty.liu1@siat.ac.cn | |
| Nan Li | Master Student | Medical Image Segmentation | nanli69-c@my.cityu.edu.hk | |
| Yuyang Liu | PhD Candidate | Computer Vision | sunshineliuyuyang@gmail.com | |
| Ziyu Liu | PhD Candidate | Medical Image Segmentation | zy.liu4@siat.ac.cn | |
| Canwei Dong | Master Student | 3-D reconstruction | cw.dong@siat.ac.cn | |
| Shoujun Zhou | Full Professor | Medical image processing | sj.zhou@siat.ac.cn | |

**Note:**

[1] This form helps us to understand your paper better; the form itself will not be published.

[2] *Prefix*: can be chosen from Master Student, PhD Candidate, Assistant Professor, Lecturer, Senior Lecturer, Associate Professor, Full Professor